\documentclass[pre,showpacs,floatfix,twocolumn]{revtex4}
\usepackage{graphicx}
\usepackage{amssymb}
\usepackage{dcolumn}
\usepackage{bm}
\begin{document}

\title{Observation of the two way shape memory effect in an atomistic model of martensitic transformation}
\author{E. A. Jagla}

\affiliation{Centro At\'omico Bariloche and Instituto Balseiro (UNCu), Comisi\'on Nacional de Energ\'{\i}a At\'omica, 
(8400) Bariloche, Argentina}

\begin{abstract}

We study a system of classical particles in two dimensions interacting through an isotropic pair potential that displays a martensitic phase transition between a triangular and a rhomboidal structure upon the change of a single parameter. Previously it was shown that this potential is able to reproduce the shape memory effect and super-elasticity, among other well known features of the phenomenology of martensites. Here we extend those previous studies and describe the development of the more subtle two-way shape memory effect. We show that in a poly-crystalline sample, the effect can be associated to the existence of retained martensite within the austenite phase. We also study the case of a single crystal sample where the effect is associated to particular orientations of the dislocations, either induced by training or by an ad hoc construction of a starting sample.

\end{abstract}
\maketitle

\section{Introduction}

Martensitic transformations are temperature driven non-diffusive phase transitions in which the atoms of a solid perform small individual atomic displacements between a high temperature high symmetry austenite (A) phase, and a low temperature lower symmetry martensite (M) phase\cite{otsuka}. When the symmetry of the martensite phase is a sub-group of that of the austenite phase, the system displays the remarkable shape memory effect (SME)\cite{bhattacharya,bhatta_nature}: The sample in the martensite phase can be mechanically deformed a big amount, yet, when the temperature is raised and the austenite becomes stable, the sample returns to its original shape. In simple terms the origin of the shape memory effect is associated to the fact that in the martensite phase, and even if the sample is mechanically deformed, each atom preserves the atomic neighborhood it had in the austenite, and when temperature is raised and austenite becomes stable again, each single atom returns to its original position.

\begin{figure}[h]
\includegraphics[width=8cm,clip=true]{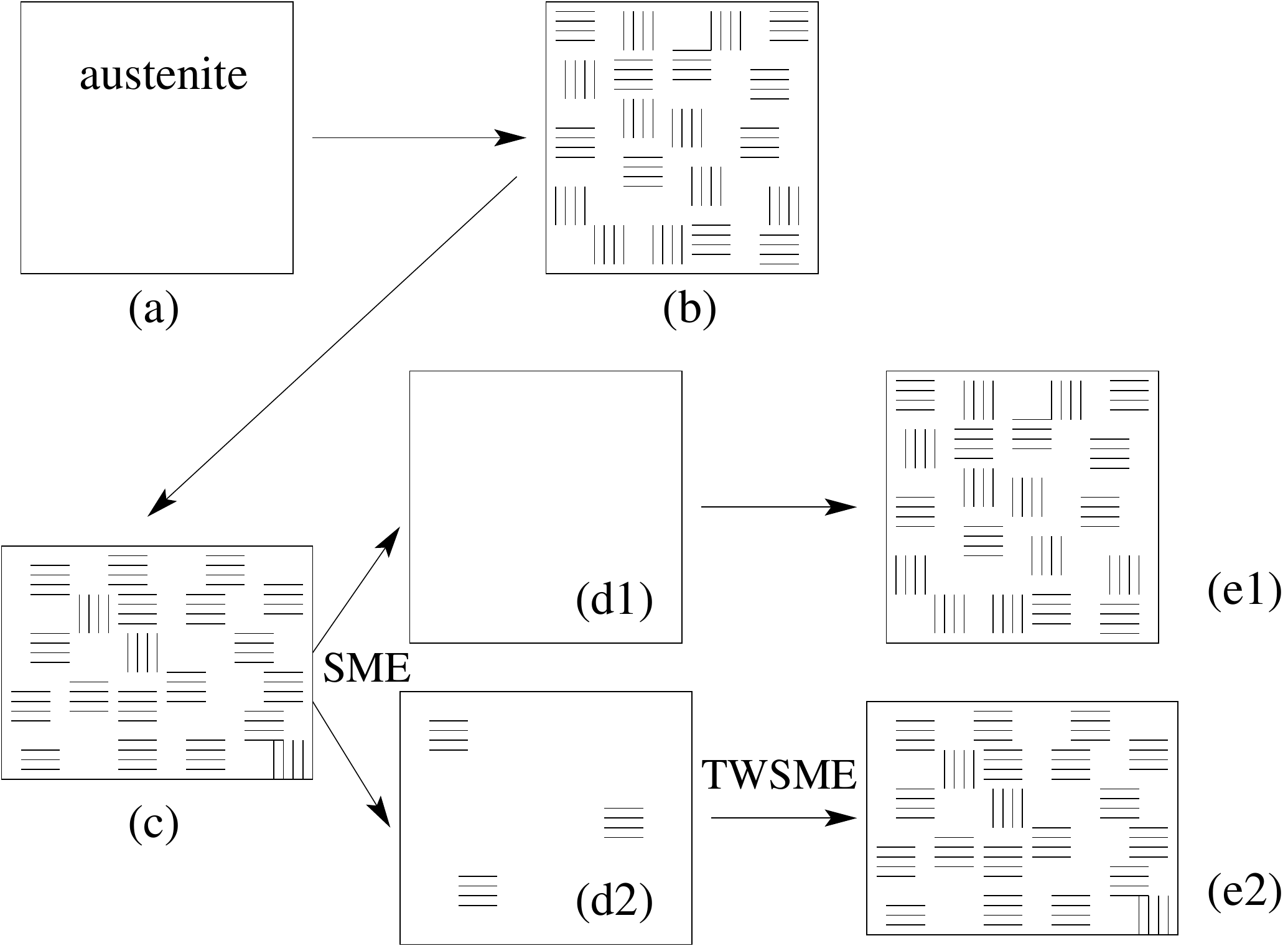}
\caption{Sketch of the shape changes in a sample displaying a martensitic transformation. Martensite is schematically indicated by the dashed regions, which can be vertical or horizontal, indicating a martensitic variant elongated in the corresponding direction. (a) Starting austenite sample. (b) Isotropic martensite after cooling. (c) Martensite deforms due to interconversion of variants, after an applied stretching along $x$. (d) Sample returns to the austenite phase after heating, recovering totally (d1) or partially (d2) its original form (the amount or remaining deformation in (d2) may be hardly visible experimentally). This is the SME. In the last case, the retained martensite in (d2) can trigger the spontaneous shape change of the sample after cooling (e2), giving rise to the TWSME.
}
\label{f1}
\end{figure}

It is quite remarkable that in many cases, after a sample was submitted to one (or more) of these cycles, it is able to ``remember" the kind of deformation it was submitted to, and upon a new cooling step, it deforms spontaneously to the remembered shape, without the application of an external deformation. The shape change under this cooling-heating cycling is termed the two way (or all-round) shape memory effect (TWSME).
The origin of the TWSME is more subtle than the simpler SME. It does not originate in general symmetry considerations, but in the remaining, after the initial cooling-deformation-heating training cycles, of 
atomic arrangements within the austenite that encode the form in which the sample was stressed, and upon a new cooling, they favor the nucleation of martensite in preferential orientations. It has been observed that the TWSME can be triggered by small remaining pieces of the martensite phase within the austenite (this is referred to as retained martensite), collections of dislocations with particular orientations, or regions of the austenite with anisotropic remnant strains.\cite{stalmans,xu,perkins,rios,cingo1,cingo2}
The generic scheme giving rise to the TWSME is qualitatively depicted in Fig. \ref{f1}.

Although well understood in qualitative terms, many details of the TWSME have not been totally elucidated.
Particularly, the thermodynamical, or dynamical reasons that favor the formation of 
retained martensite or other kind of elastic distorsions that trigger the effect are only poorly understood.
Also, other interesting questions can be stated, as for instance to what extent a starting austenite sample can be engineered, in such a way to obtain a predefined deformation effect upon cooling, without the necessity of training cycles.

In this paper we contribute to answer these questions. In section II we present an atomistic two-dimensional model that displays a martensitic phase transition between a triangular and a rhomboidal phase. In section III we construct a polycrystalline sample with periodic boundary conditions, and apply a 
cooling-deformation-heating training cycle. After that, we investigate if the TWSME appears. We observe the appearance of the TWSME for the case in which the deformation during the training cycle is large enough, which is correlated with only a partial recovery of shape after heating in the training cycle, and with the existence of preferential orientations of retained martensite in the austenite phase.  In section IV we show the same effect for the more realistic situation of a long bar upon bending. In Section V we study the TWSME in a single crystal sample with dislocations, showing how preferential orientations of the dislocations can justify the appearance of the TWSME. We also
address the potentially promising possibility of designing a sample with an ad hoc distribution of defects that displays the TWSME. We show that a particular spatial distribution of dislocations can be proposed for which a robust and persistent TWSME is obtained.
Finally, in Section VI we summarize and conclude.

\section{Model}

Martensitic transformations typically occur due to an {\em entropic stabilization} mechanism of the austenite phase\cite{elliot}. The bare interaction potential between the particles has the deepest minimum at the structure of the martensite phase, whereas the austenite phase is at most a secondary metastable minimum. Yet, the basin of attraction is wider for the austenite than for the martensite, so that thermal fluctuations generate a decrease of the free energy of the austenite with respect to the martensite as temperature is raised. At a certain temperature the two structures can exchange stability, and the transition occurs.

A full numerical simulation of a martensitic transformation requires the simulation of the dynamics during a long time, to observe the transition between the structures taking place. In addition, the elementary time step of the simulation must be small enough in order to reproduce the thermal movement of the particles. On the whole, this fully first principle simulation requires a large temporal span that makes the simulation very time consuming. Simulations along these lines have in fact being done\cite{kastner1,kastner2,kastner3}, however the subtle effects we want to detect are hard to observe with this technique.

In an alternative approach, we consider an effective interaction potential that already includes the effects of entropy. In this sense, this atomic potential must be actually considered to be a free energy functional in the spirit of Ginzburg-Landau free energies used in general descriptions of phase transitions. We will assume that the effect of temperature is to change the form of this effective free energy potential. 
The stability range between martensite and austenite structures is directly encoded in the form of the potential, and then the simulation will display the transition without the need to truly simulate the atomic thermal vibrations.

We numerically study the system by solving for the time dependence of the particle coordinates
according to a standard Verlet scheme in the presence of a frictional term proportional to velocity. This term takes out of the system any kinetic energy generated during the transformation, and effectively keeps the system at a local energy minimum.

The potential we use and its basic features were presented in \cite{laguna}.
For completeness we give the form of the potential here. The interparticle potential $V(r)$ is composed of several parts:
\begin{equation}
V(r)=V_0+V_1+V_2+V_3
\end{equation}
where

\begin{eqnarray}
V_0&=&A_0 [1/r^{12}-2/r^6+1]~{\mbox if}~~r<1\\
V_1&=&[(r-1)^2(r+1-2c)^2 / (c-1)^4] - 1\\
~&~&~~~~~~~~~~~~~~~~~~~~~~~~~{\mbox if}~~r<c\nonumber\\
V_2&=& - A_2 [(r-d_2-s_2)^2 (r-d_2+s_2)^2] / s_2^4\\
~&~&~~~~~~~~~~~~~~~~~~~~~~~~~{\mbox if}~~d_2-s_2<r< d_2+s_2\nonumber\\
V_3&=& A_3 [(r-d_3-s_3)^2 (r-d_3+s_3)^2 ] / s_3^4\\
~&~&~~~~~~~~~~~~~~~~~~~~~~~~~{\mbox if}~~d_3-s_3<r< d_3+s_3\nonumber
\end{eqnarray}
and $0$ otherwise in all cases.
$V_0$ is the repulsive part of a LJ potential and its weight in the total potential is measured by the parameter $A_0$.  The fourth order term proportional to $V_1$ contributes with an attractive well to the total potential. The last two terms are  fine tuning terms that provide a small  minimum of amplitude $A_2$ centered at $d_2$, and a small maximum of amplitude $A_3$ centered at $d_3$. They were adjusted to penalize appropriately the triangular lattice, and/or favoring the martensitically related structure. The potential is fully determined by the set of parameters $P=\{A_0, A_2, A_3, c, d_2, s_2,  d_3, s_3\}$.
To study the triangular-rhomboidal transition (T-R) we use
$P =\{A_0, 0.003, 0.01, 1.722, 0.98, 0.04, 1.74, 0.2\}$ with variable $A_0$. $A_0$ is the parameter that models the effect of temperature. There is a critical value of $A_0^c\simeq 0.067$ such that for $A_0 < A_0^c$
($A_0>A_0^c$) the martensite  (austenite) phase is globally more stable.

The lattice parameter of the austenite phase is approximately 0.995 at the transition point, increasing slightly with the value of $A_0$. The austenite phase is characterized by two different interatomic distances that are approximately 0.93 and 1.04 at the transition. In order to visualize parts of the sample in the austenite or martensite phase, and since a direct examination of every atomic position is impractical, we make the following analysis. Given a configuration of the particles, neighbor particles located at a distance compatible with the largest interatomic distance in the martensite phase  are identified (for practical reasons we use a distance window between 1.015 and 1.2), and those links are plotted as segments. This highlights regions of the sample in the martensitic phase. In addition, since our attention is focused on the shape of the sample, the following color code is implemented. When the angle between the long segment of the martensite and the $x$ axis is less than 45 degrees, the segmant will be plotted in red, whereas if this angle is larger than 45 degress, it will be plotted as green. This color coding allows to identify at a glance if the sample (or a part of it) is preferentially stretched in the $x$ or $y$ direction.

\section{Shape memory and two-way shape memory in a periodic, polycrystalline sample}

The ``shape" of a sample is a characteristic that is strongly connected to the existence of free surfaces. 
Yet as a first example, and since we are looking for tiny effects that must show up in the simulations, a different set up will be more convenient. 
We will study in this section a system with periodic boundary conditions within a rectangular box of size $l_x$, $l_y$.
In this set up, the shape of the sample is characterized by the ratio $l_x/l_y$. The values of $l_x$ of $l_y$ are 
allowed to adjust dynamically during the simulation according to the following mechanism. The internal compressions $f_x$ and $f_y$ in the system in both direction $x$ and $y$ are calculated along the simulation
and are driven towards externally imposed values$f^{ext}_x$ and $f^{ext}_y$, 
by the adjustment of $l_x$ of $l_y$, according to the equations

\begin{equation}
\frac{dl_{x,y}}{dt}=\lambda (f_{x,y}-f^{ext}_{x,y})
\end{equation}
If $f^{ext}_{x,y}$ are set to zero, the equilibrium values of $l_{x,y}$ determine the shape of a ``free" sample. 
Instead, non-zero values of $f^{ext}_{x,y}$ model the application of external forces along the $x$ or $y$ directions.

\begin{figure}[h]
\includegraphics[width=8cm,clip=true]{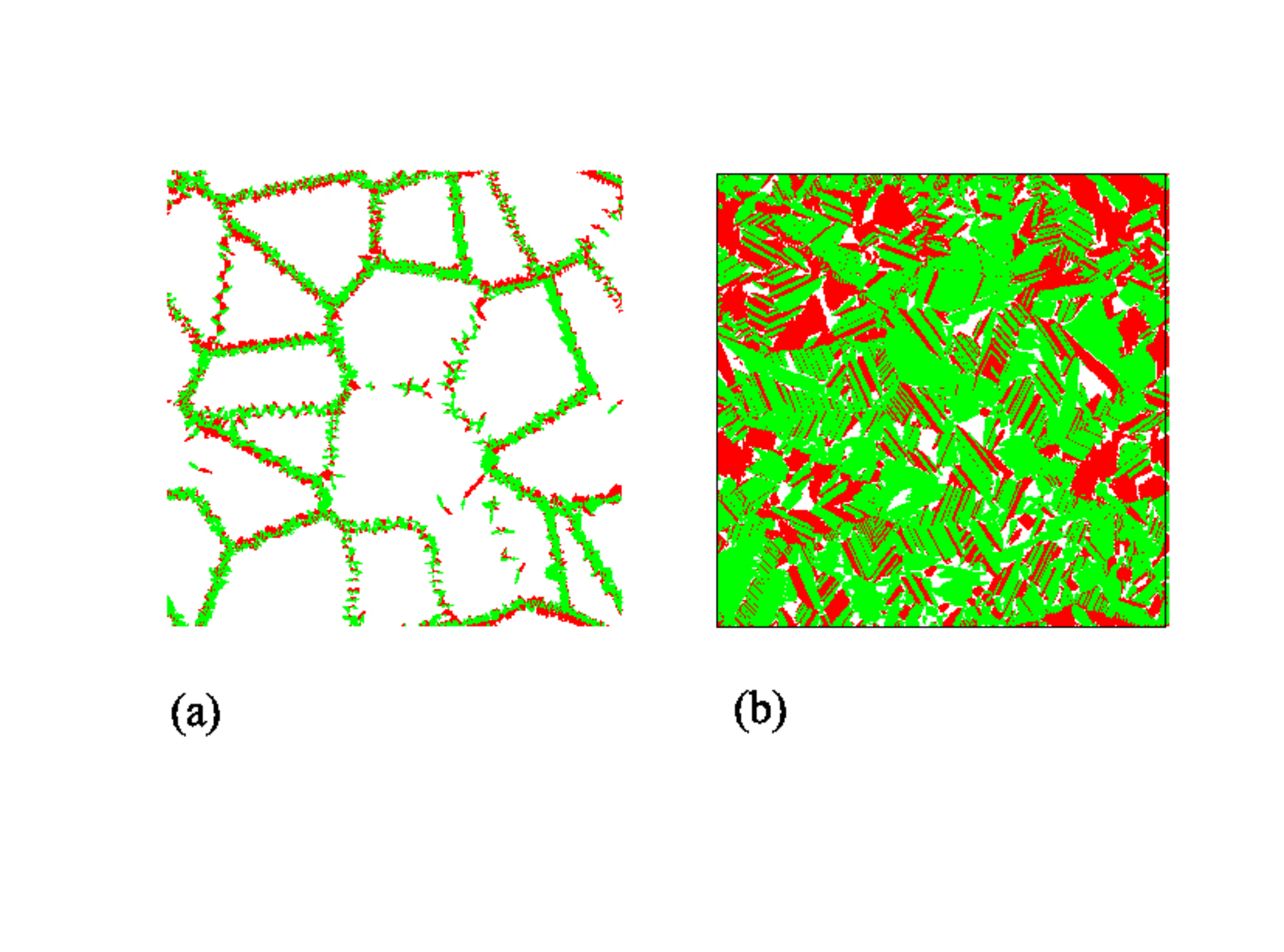}
\caption{(a) The starting polycrystalline sample after initial relaxation in the austenite phase. Plotting details and the meaning of the color code are provided in the text. Note that although the bulk of the grains are in the austenite phase, atoms with interatomic distances compatible with the martensite are widely present at the grain boundaries. (b) The sample after transformation to the martensite phase. There is no appreciable shape change during this transformation. 
}
\label{periodico1}
\end{figure}

The starting sample consists of a poly-crystal constructed in the following way. A number of grain centers and orientations are randomly chosen, and they are used to generate a given grain orientation around each of the centers. Each grain extends in the sample according to a Voronoi tessellation criterion, namely each sector of the sample is dominated by the nearest grain center.
As a first step, this initial configuration is relaxed with the true interatomic potential and $A_0=0.085$, namely well within the stability range of the austenite phase. The values of $l_{x,y}$ are also allowed to relax during the process, under $f^{ext}_{x,y}=0$. The configuration obtained is shown in Fig. \ref{periodico1}(a). We observe that although the grains are in the austenite phase in bulk, in the grain boundaries there are atoms that are at the appropriate relative distance so as to trigger the martensite transformation.
As $A_0$ is reduced and the martensite becomes progressively more stable, martensite crystals grow from the grain boundaries and invade the bulk of the grains, until all the sample is transformed [fig. \ref{periodico1}(b)]. In this transformation the form of the sample does not change, beyond some fluctuations associated to finite size effects.

\begin{figure}[h]
\includegraphics[width=8cm,clip=true]{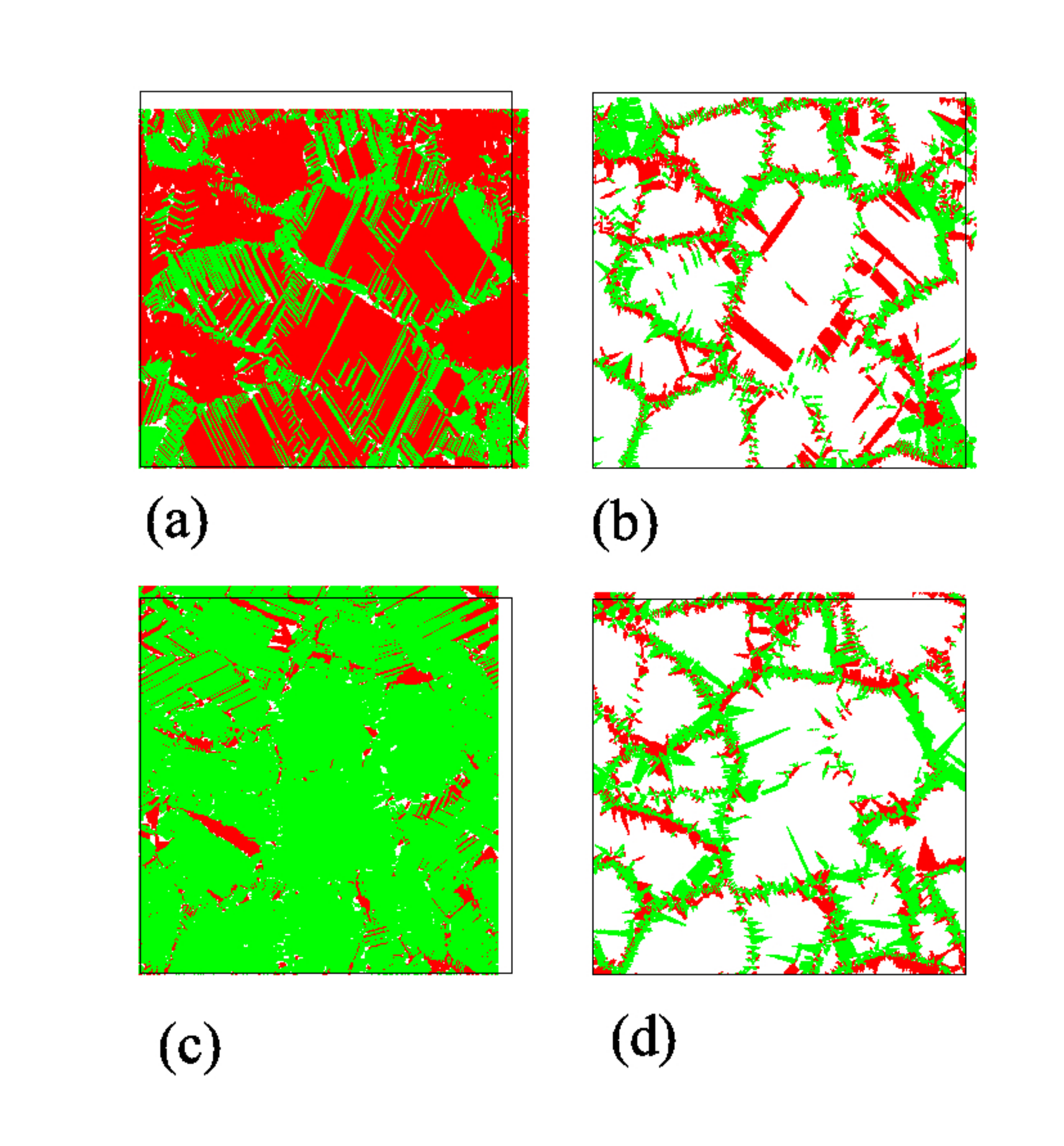}
\caption{(a) The sample in Fig. \ref{periodico1}(b) after external stretching in the $x$ direction and 
further relaxation once the stretching is withdrawn. (b) State after re-transformation to the austenite. (c) and (d) The same for an initial compression along $x$ direction.
The black square is an outline of the sample in Fig. \ref{periodico1}(b), for comparison.
Note that the remaining martensite in (b) and (d) corresponds to the deformation produced by the applied stretching.
}
\label{periodico2}
\end{figure}

 At this stage we apply an external deformation through the setting of non-zero values of $f^{ext}_x=-f^{ext}_y$, until we get a prescribed change in the shape of the sample, characterized by the ratio $l_x/l_y$. Fig. \ref{periodico2}(a) depicts the structure of the sample after a $9\%$ change in $l_x/l_y$, and ulterior relieve of the external force and relaxation. Interconversion of variants caused by the external deformation (favoring $x$-elongated red ones) is apparent, and the sample retains a deformation of about $5\%$. From this configuration, the value of $A_0$ is taken back to 0.085, 
and the final configuration is shown in Fig. \ref{periodico2}(b). Note that although the sample mostly returns to its initial configuration, displaying SME, there are remains of martensitic variants, mostly red ones, that correspond to pieces of martensite compatible with the previously applied deformation. This produces that the shape of the sample in Fig. \ref{periodico2}(b) is not exactly equal to that in Fig. \ref{periodico1}(b), but some elongation along $x$ remains. Fig. \ref{periodico2}(c) and (d) show the equivalent situation that occurs when the external stretching is performed along the $y$ direction. In this case the remaining martensite in Fig. \ref{periodico2}(d) is mostly of the green type, and the sample retains an elongation along the $y$ direction. 

\begin{figure}[h]
\includegraphics[width=8cm,clip=true]{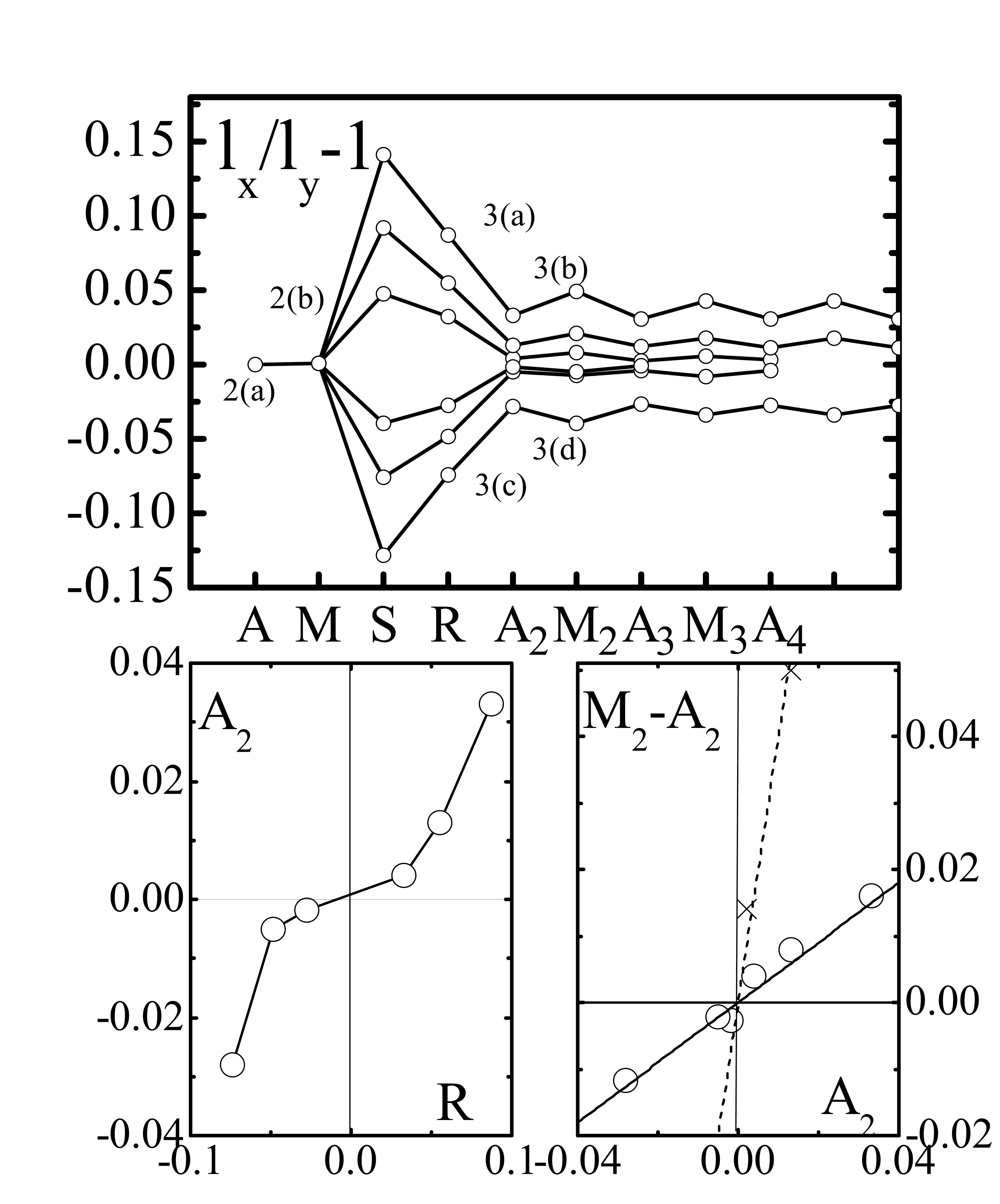}
\caption{(a) Evolution of the shape ($l_x/l_y$) of the sample according to the following protocol. $A$: starting austenite (Fig. \ref{periodico1}(a)). $M$: transformation to the martensite (\ref{periodico1}(b)). $S$: stretched sample upon application of different external loads. $R$: relaxed sample after external force is set to zero (examples in \ref{periodico2}(a),(c)). $A_2$: transformation to the austenite, and observation of the SME. $M_2$, $S_2$ and the following: shape change under cyclic austenite-martensite transformation, in the absence of external loading. This is the TWSME.
(b) The remnant deformation as a function of the maximum deformation. The form of this curve is clearly non-linear. (c) The amount of shape recovery  in the TWSME as a function of the remnant deformation in the austenite phase. This relation is seen to be almost linear in the range of deformations analyzed, with a proportionality factor of about 0.5. The two crosses and the dotted line show a few results in a sample with larger grain sizes, where this factor is much bigger.
}
\label{forma1}
\end{figure}

The detailed evolution of the shape of the sample, as measured by $l_x/l_y$ is presented in Fig. \ref{forma1}(a): The shape does not change in the first $A$$\to$$M$ transformation. The sample elongates upon the application of the external stress ($S$). When the stress is relieved ($R$), the deformation is reduced, but the sample remains deformed. When the sample returns to the austenite phase ($A_2$)
it mostly recovers its original shape. This is the standard SME. 
From the configurations $A_2$, we perform new austenite-martensite cycling, without applying any external stress. 
The evolution of the shape during this process is seen in Fig. \ref{forma1} ($A_2\to M_2 \to A_3$...). We see that a systematic TWSME is observed, that follows the training imposed by the original external deformation. 
To better quantify this effect, we first plot in Fig. \ref{forma1}(b) the amount of remnant deformation after the first re-transformation to the austenite ($A_2$) as a function of the deformation caused by the external loading ($R$). We see that this dependence is strongly non-linear. It is vanishingly small if the amount of applied deformation is low, but it rapidly increases at larger applied deformation. In (c) we see that the intensity of the TWSME is proportional to the remaining deformation after the training step indicating that preferential orientation of the remaining martensite within the austenite is the main responsible of TWSME. In fact, the remaining deformation in the austenite can be argued to be proportional to the imbalance between the different orientations of the retained martensite. In turn, this imbalance is responsible for the preferential growth of conveniently oriented variants when the sample transforms to the martensite phase.

It is worth to be mentioned that the TWSME we observe is robust with respect to cycling between martensite and austenite.  Typically the effect is maximum in the first cycle, it diminishes  about 20$\%$ in the second cycle, and then it conserves this value for at least 20 cycles, that was the longest run we performed.

In experimental realizations,  it has been observed that the TWSME can be almost perfect: the sample may recover the deformed shape, even if the remnant deformation is so small that it is undetectable. 
On the contrary, the plot in Fig. \ref{forma1}(c) shows that the amount of the TWSME in our case is tiny: it approximately corresponds to a spontaneous deformation upon transition to the martensite that is about one half of the remnant deformation in the austenite after training. We think that this is related to the grain size of our sample. In fact, in additional simulations using smaller grains, the TWSME was almost unobservable. Instead, in a more demanding simulation using grains of linear size about three times those in Fig. \ref{periodico1}, the effect was much stronger, as the two crosses in Fig. \ref{forma1}(c) indicate. The dotted line indicates in this case that the shape recovery is about five times larger than the remnant deformations. We think that this trend makes plausible that the TWSME we are observing is compatible with that observed experimentally, where the grain sizes are much larger than those in our samples.

\section{SME and TWSME in a free bar}

The results in the previous section show clearly that our model contains all the necessary ingredients to explain the origin of the TWSME. It would be desirable however, to show the effect appearing in an experimentally achievable configuration, particularly in a sample with free surfaces. 

\begin{figure}[h]
\includegraphics[width=8cm,clip=true]{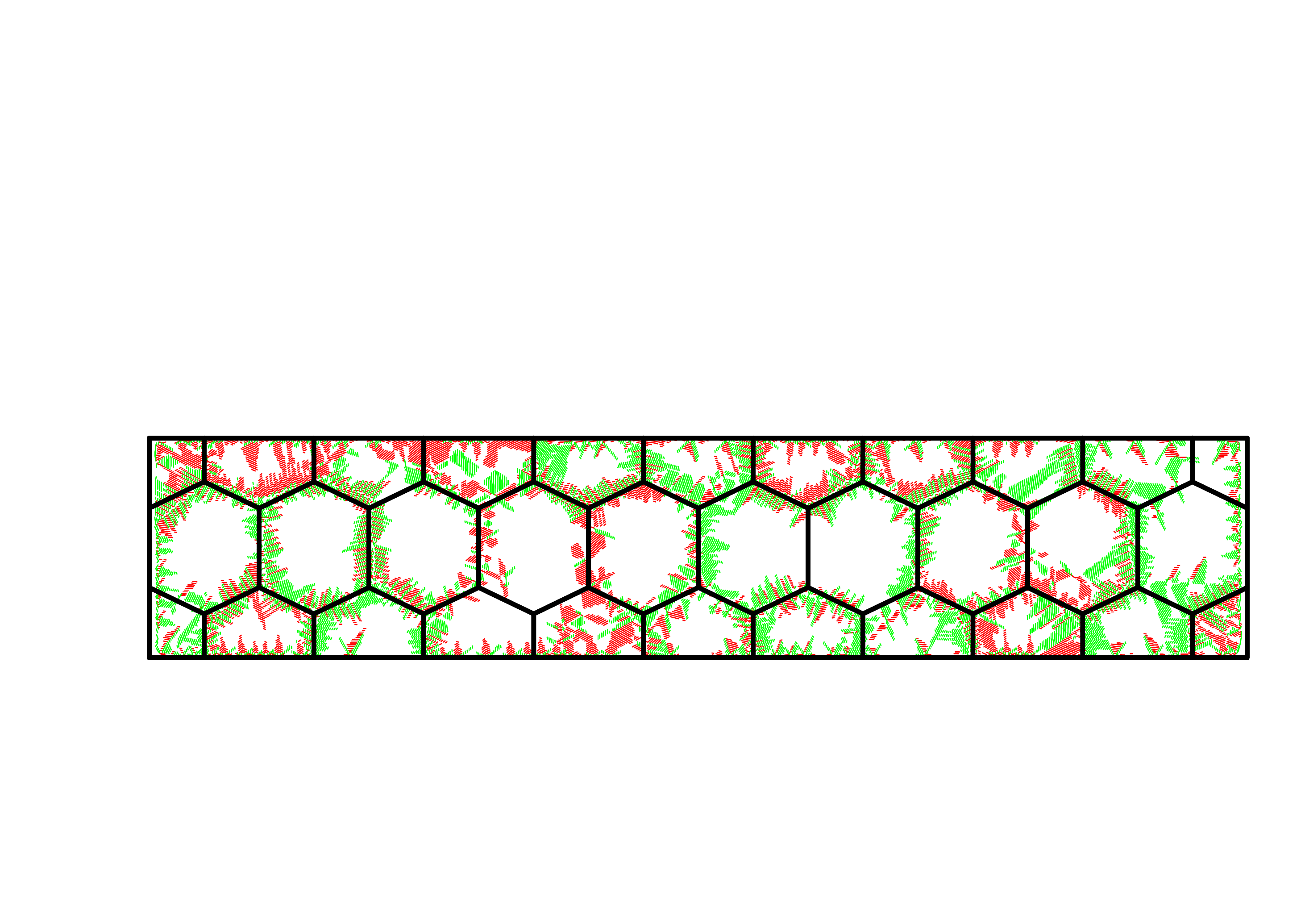}
\caption{The bar shaped sample. The black lines outline the grains in the starting sample. The actual configuration of the sample after initial relaxation in the austenite phase is shown (the crystalline orientation of each grain was chosen randomly). }
\label{barra1}
\end{figure}

\begin{figure}[h]
\includegraphics[width=8cm,clip=true]{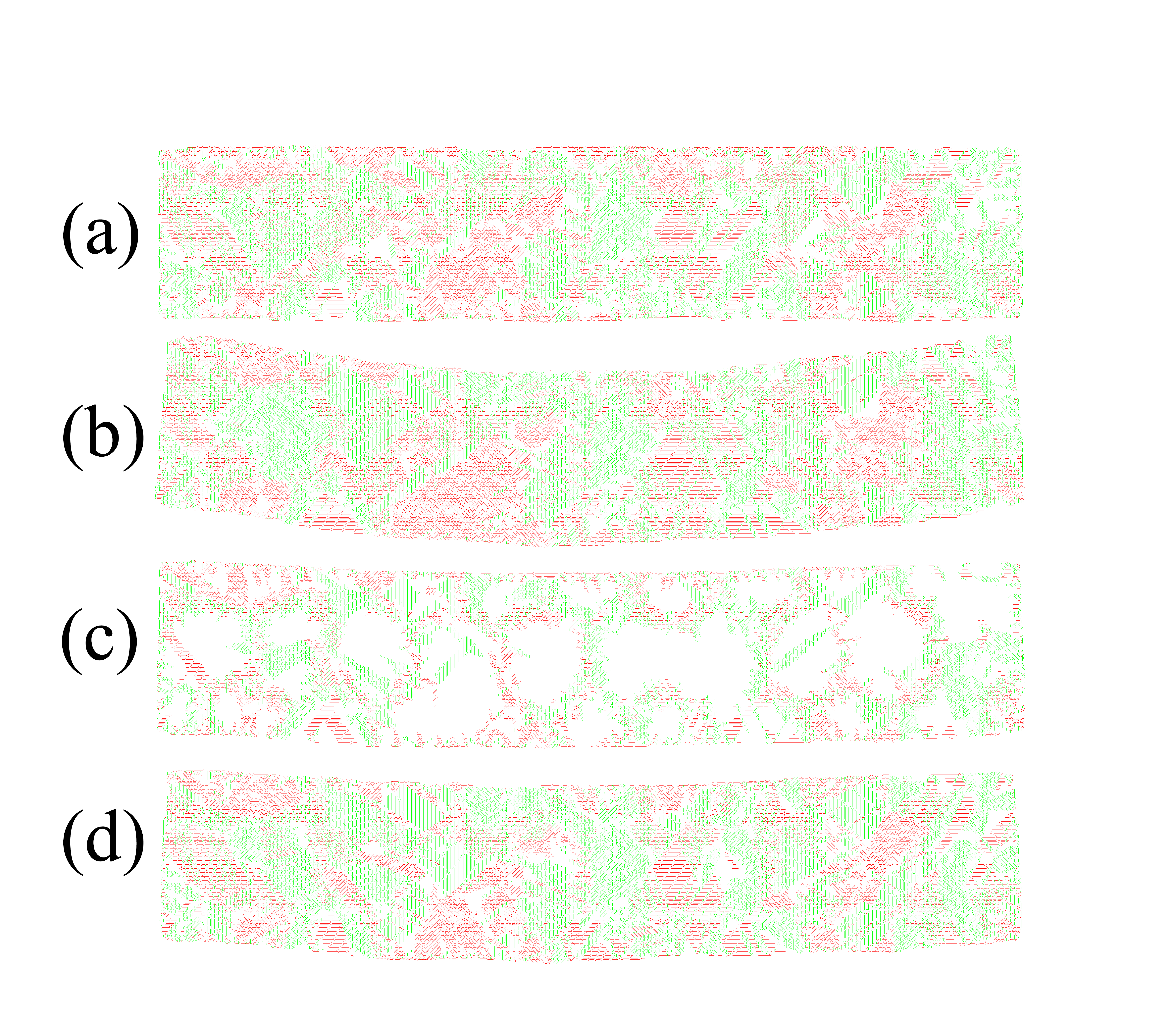}
\caption{Evolution of the bar along the treatment. (a) Free sample in the martensite state. (b) Bending caused by an external loading, after the loading has been relieved. (c) Transformation to austenite. The sample straights, showing SME, but an appreciable deformation remains. (d) Re-transformation to the martensite phase. The sample bends (a small amount) spontaneously. This is the TWSME.}
\label{barra2}
\end{figure}

In order to do this, we consider the case of the bending of a bar shaped sample, with free surfaces. 
Bending is produced in the simulations by the application of external forces along the vertical ($y$) direction onto every particle, with a magnitude $F_y$ that depends on the $x$ coordinate of the particle as $F_y=F_0((2x/L)^2-1/3)$ (the bar is located between $x=-L/2$ and $x=L/2$), and $F_0$ is chosen to achieve the desired degree of deformation.

The internal structure of the sample is again poly-crystalline, but in this case the spatial distribution of the grains was chosen by hand. 
The reason for this choice is that randomly placed grains produce too strong variations from sample to sample in our small systems.
The outline of the grains, and the actual initial configuration after relaxation are shown in Fig. \ref{barra1}. The system is submitted to a sequence of transformations as displayed in Fig. \ref{barra2}. First: reduce $A_0$ down to 0.057 to reach the martensitic state (a). In this process the shape of the sample does not change appreciably. Second: application of an external bending stress followed by the removal of the stress. 
The final configuration obtained is shown in (b), where it is seen that an important bending deformation remains in the sample. Third: transform back to the austenite, by increasing $A_0$ to 0.085 (c), in this stage the SME shows up, and the bar tends to recover its original straight shape. Note however that the recovery is not complete: a certain amount of deformation remains that is originated in pieces of $x$-elongated (green) martensite in the lower part of the sample, and $y$-elongated (red) martensite in the upper part. This imbalance of martensitic variants are the germ of the spontaneous bending of the sample when we transform to the martensite phase ($A_0=0.057$) in the absence of external stresses. The final configuration is shown in (d). 

\begin{figure}[h]
\includegraphics[width=8cm,clip=true]{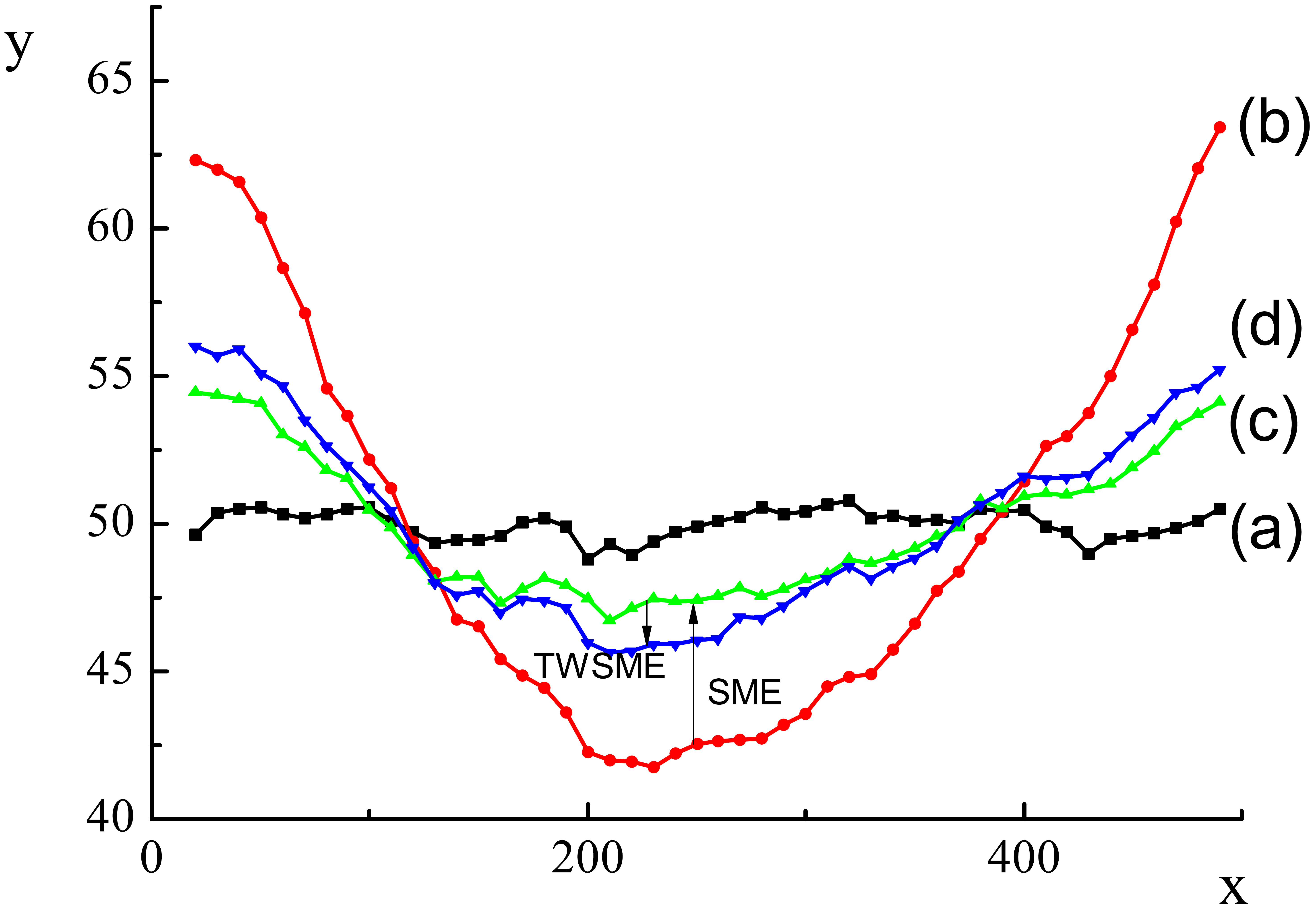}
\caption{Profiles of the samples in the previous figure, to quantify the degree of bending in each stage.}
\label{fibra}
\end{figure}

To quantify the TWSME, which is tiny for our small samples, we calculated the average vertical coordinate of the particles in  slices of the bar at different horizontal coordinates, and these values are plotted in Fig. \ref{fibra} for the configurations in Fig. \ref{barra2}. We see that in addition to the clear SME between (b) and (c), a small but clearly noticeable TWSME between (c) and (d) exists, in which the sample bends spontaneously without the application of any stress. This effect is persistent upon successive cycling between austenite and martensite, and it is of the same intensity as the one described in the previous section for a sample with periodic boundary conditions. In fact, quantifying the effect by the curvature of the bar, we see that the shape recovery when the sample transform to the martensite (the ``TWSME" arrow in Fig. \ref{fibra}) is about 0.4-0.5 of the remnant deformation (the (c) profile in the same figure), very much as in Fig. \ref{forma1}(c).

\section{SME and TWSME in single crystals with dislocations}

In the previous Sections, the origin of the TWSME was related to the remaining of martensite variants within the austenite structure. For the systems analyzed, these remains were favored by the poly-crystal nature of the sample, that allows for low energy atomic rearrangements at the grain boundaries.

Experimentally, TWSME is known to occur also in the case of single crystal samples. In fact, although a perfect austenite single crystal is not able to encode within its structure a memory of a previous deformation in the martensite phase, the unavoidable existence of defects --particularly dislocations-- in the crystal structure provides a mechanism to generate the TWSME. 
Dislocations are defects that produce anisotropic stresses in the crystal structure, which can favor the growth of martensite variants of particular orientations upon cooling \cite{stalmans,sade,lovey}. 

Experimentally, it is known that the TWSME can be induced in single crystals by training, this means that dislocations acquire a convenient distribution to trigger the shape transformation in some prescribed way, after the removal of the external stress.
It is also possible to consider the case of a sample that is engineered with a collection of dislocations distributed in some appropriate manner, such that the sample may have a tendency to spontaneously elongate in one direction and contract in the perpendicular one upon the martensitic transformation. Although at present to produce such an ad hoc distribution of dislocations is not feasible, this seems an interesting theoretical situation to consider.

\begin{figure}[h]
\includegraphics[width=8cm,clip=true]{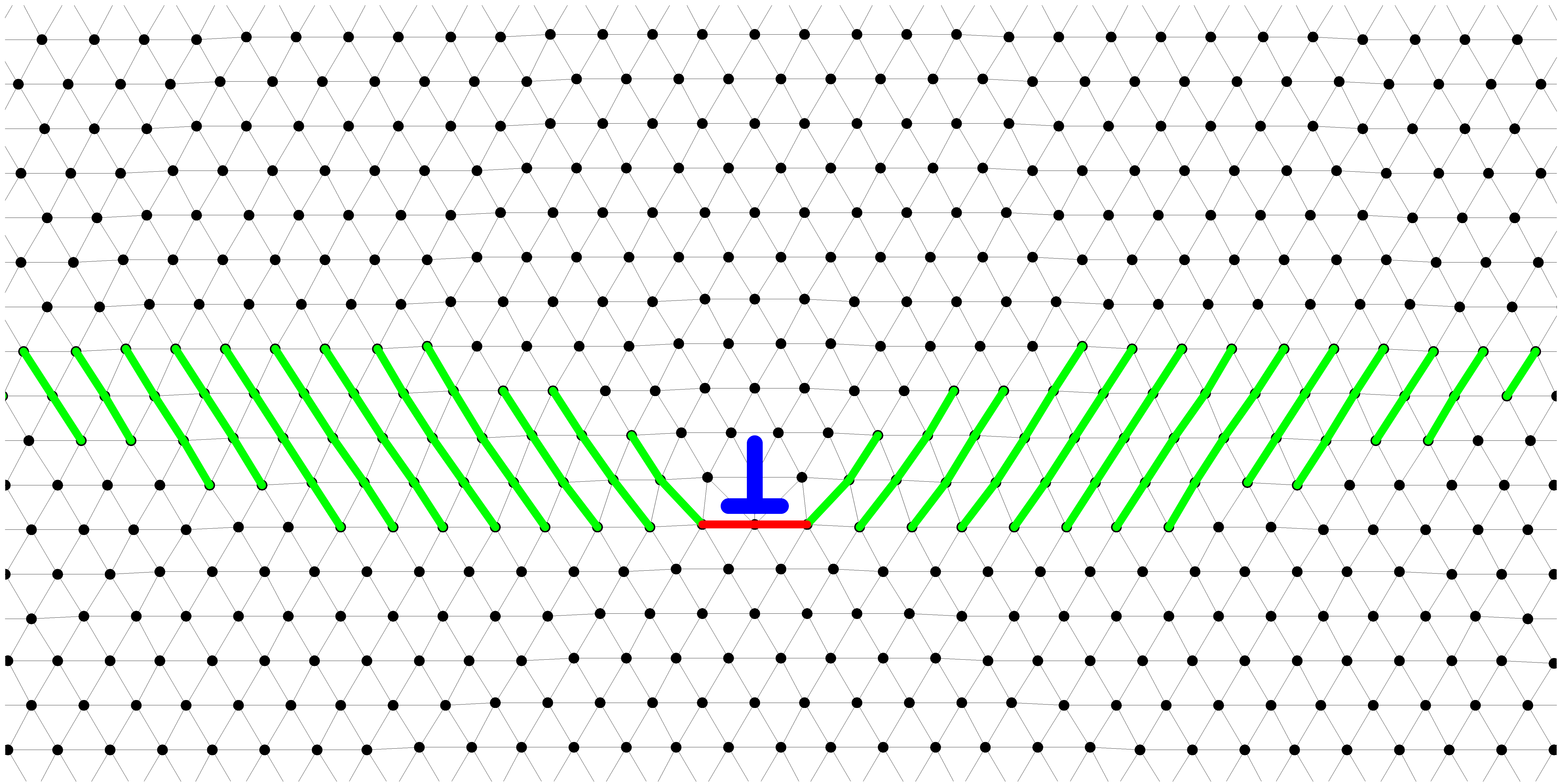}
\caption{Lattice structure around a single dislocation in a perfect triangular structure. The dislocation core is indicated by the ''$\perp$" symbol. Highlighted interatomic links indicate those that are in the range $[1.015,1.2]$, well different than the equilibrium lattice parameter of the triangular structure ($0.995$).}
\label{1dislocacion}
\end{figure}

The elastic distortions that a single dislocation generates on an isotropic elastic solid are well known. However, it is not necessary to consider this deformation in full detail here.
The following analysis is more appropriate to our purposes.
We constructed by hand a single crystal structure with a dislocation in an otherwise perfect single crystal sample. 
After preparation, the structure was relaxed using the interatomic potential in the austenite phase ($A_0=0.085$).
The final structure is shown in Fig. \ref{1dislocacion}.
As in previous figures, we also plot in that figure the interatomic links that are in the range appropriate to become the largest lattice parameter of the martensite structure, i.e., they will be the germs on which martensite crystals will grow. 
It is thus seen in the particular case of Fig. \ref{1dislocacion} that this dislocation will induce a tendency to elongate the sample along the insertion line of the dislocation ($y$ direction in this case) and a tendency to contract along the glide direction ($x$ direction).

\begin{figure}[h]
\includegraphics[width=8cm,clip=true]{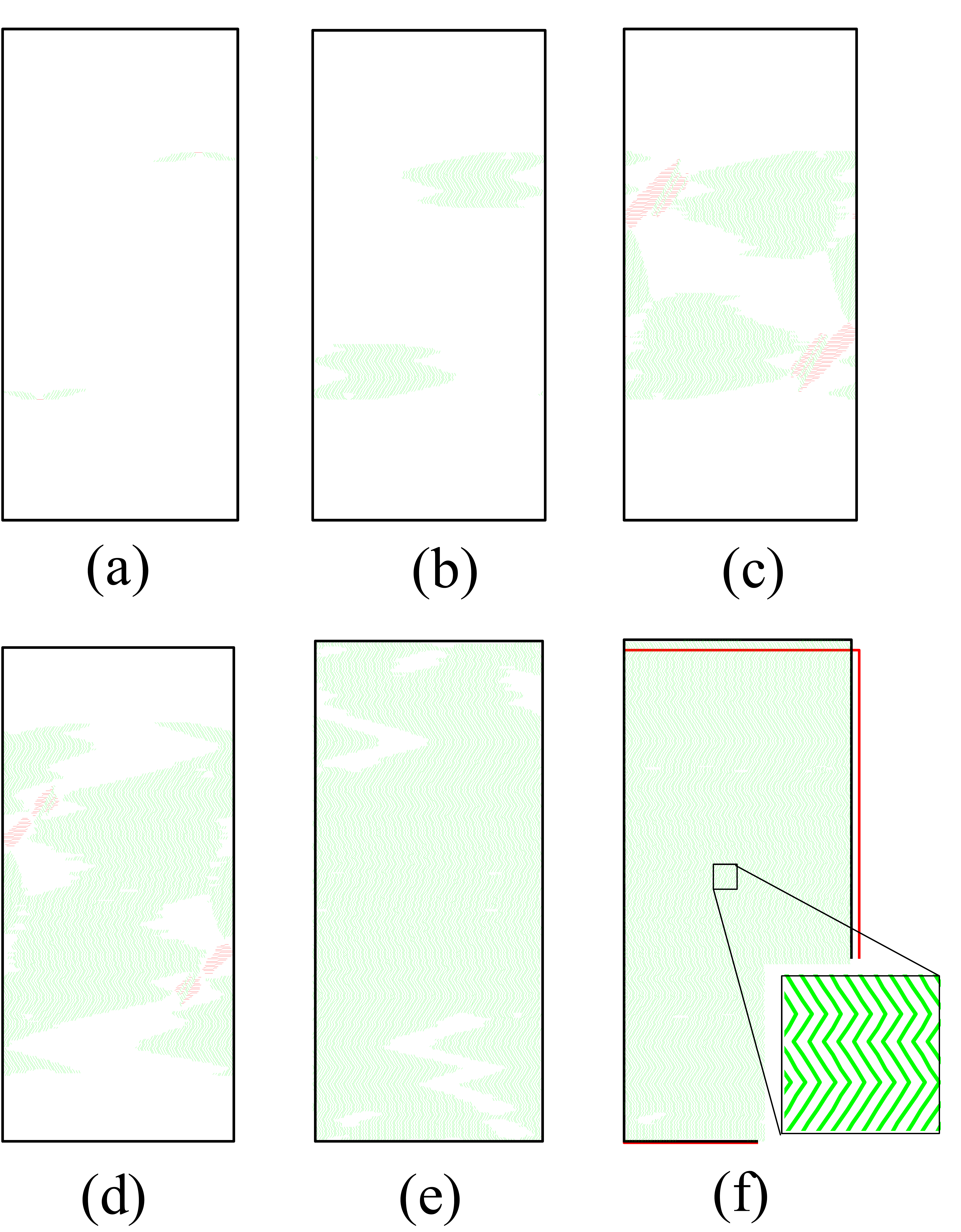}
\caption{Martensitic transition of a sample with two dislocations and periodic boundary conditions. (a) Stable configuration for $A_0=0.060$. (b) to (e) spontaneous transformation for $A_0=0.059$. (f) final configuration at $A_0=0.057$. The red rectangle is the shape of the original sample in (a), and allows to see that there was an important shape change. Note that individual links between particles are not discernible in the main plots. The zoom in (f) shows a detail of the structure.}
\label{2disloc}
\end{figure}

In order to see to what extent dislocations are effective to encode the TWSME, we first consider a simple case. To avoid the existence of free surfaces that can blur the effect of the dislocations, a single crystal with periodic boundary conditions is necessary. However, it is not possible to accommodate a single dislocation with these boundary conditions. Instead, we used a crystal with two dislocations as sketched in figure \ref{2disloc}(a). Note that both dislocations have the same orientation of the glide direction, thus contributing in the same way to a possible TWSME. The separation between the two glide lines was chosen by hand, however, note that the relative position of the two dislocation along this direction (the $x$ separation of the dislocation cores) is adjusted by the system, as dislocations are very mobile along $x$. As the $A_0$ parameter of the potential is reduced, the system remains within the austenite phase up to $A_0\simeq 0.060$ (Fig. \ref{2disloc}(a)), well below the equilibrium transition  value which is 0.067. This indicates that there is an appreciable energy barrier to the nucleation of the martensite in the present case. Reducing $A_0$ further produces an abrupt transition to the martensite phase (\ref{2disloc}(b)-(e)), in which only two of the three possible martensitic variants are present. These variants are precisely those favored by the dislocations in the system, and generate a global contraction along $x$ and expansion along $y$, as indicated in the last panel of Fig. \ref{2disloc} by the outline of the initial and final shapes. The amount of the TWSME is in this case very close to the ideal value that can be expected in this case (about 7 \% in $l_x/l_y$)
This is a clear evidence that the sample displays a strong TWSME.

The previous example shows --in a rather ideal situation-- that the TWSME may be induced in our model by dislocations. 
Looking for the same effect in  a more realistic set up, 
we tried to train a single crystal to display TWSME.
Since some amount of disorder in the starting sample is necessary, we started with a sample with a small concentration of vacancies, and performed austenite-martensite cycles in the presence of an external stress. It was observed that the vacancies reacommodated in the form of dislocations, that were distributed in the sample in a more or less random fashion. 
However, we did not observe an over abundance of dislocations with the insertion line parallel to the stretching direction, that would favor the TWSME. In fact this was not observed beyond the limit of sample to sample variations. It is not clear to us why this attempt was unsuccessful.

\begin{figure}[h]
\includegraphics[width=10cm,clip=true]{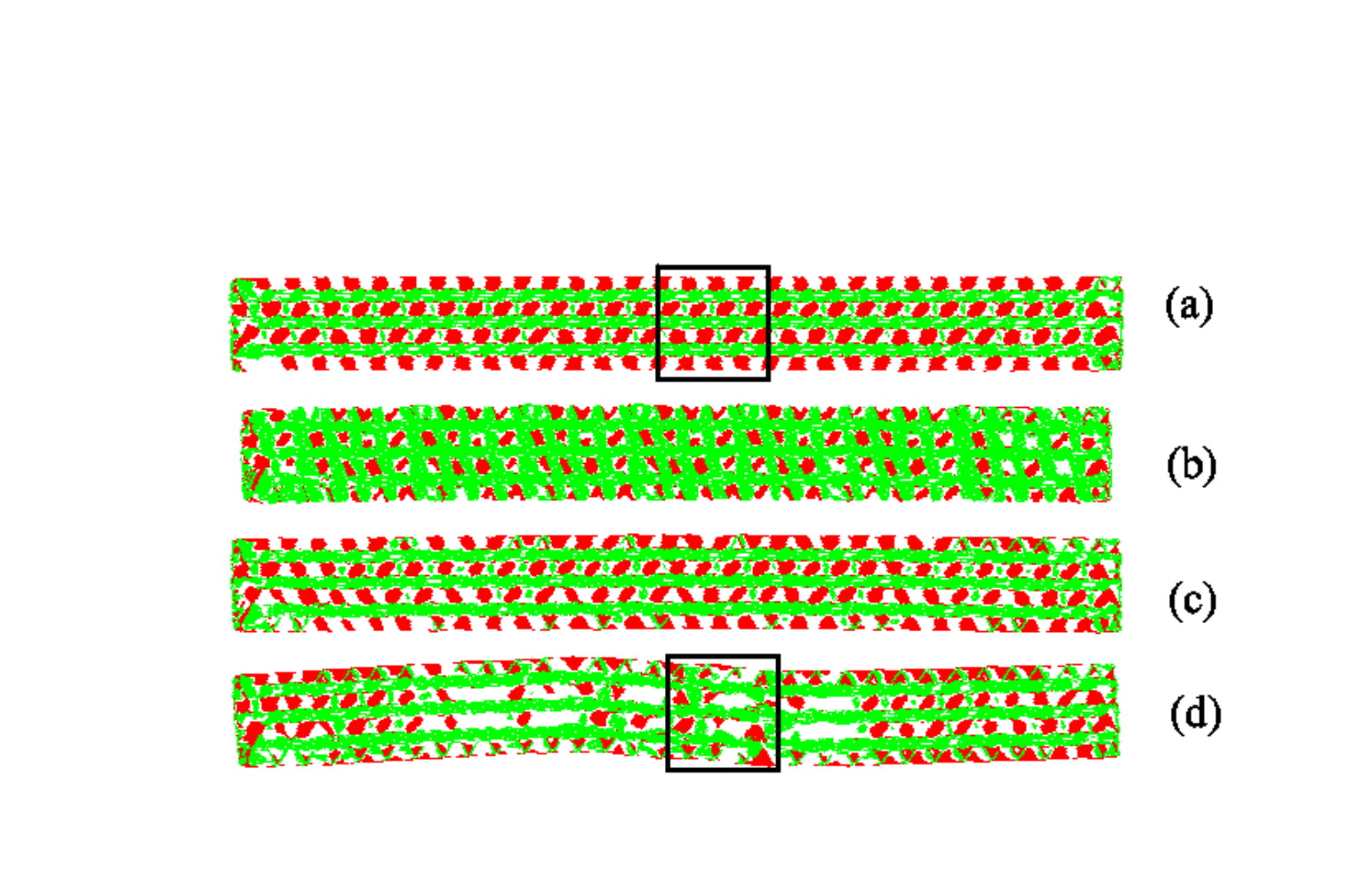}
\caption{A free bar engineered with a collection of dislocations to induce the TWSME. (a) Structure after a relaxation with a parameter $A_0$ corresponding to the austenite phase. (b) The sample after the first transition to the martensitic, and (c) back to the austenite.
(d) The austenite sample after a few cycles, where some degradation is observed. Details of the configurations in the two outlined boxes are shown in the next figure.}
\label{many_disloc}
\end{figure}

\begin{figure}[h]
\includegraphics[width=8cm,clip=true]{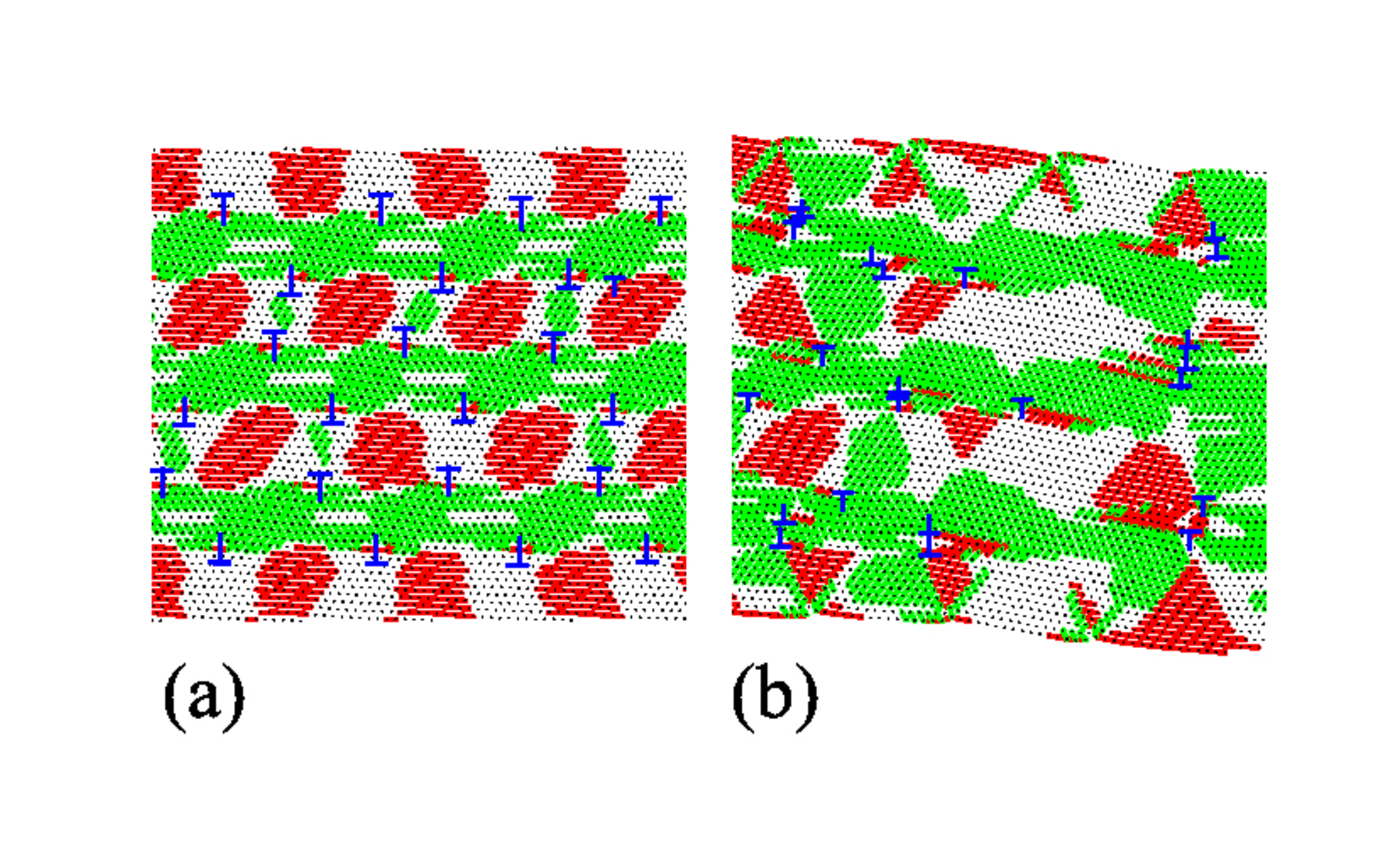}
\caption{Detail of the configurations in the two square boxes highlighted in the previous figure. In (a), note the almost periodic arrange of dislocations that were introduce by hand. In (b) we show a particular part of the sample that has suffered of an important degradation due to dislocation movement.}
\label{perp}
\end{figure}

\begin{figure}[h]
\includegraphics[width=8cm,clip=true]{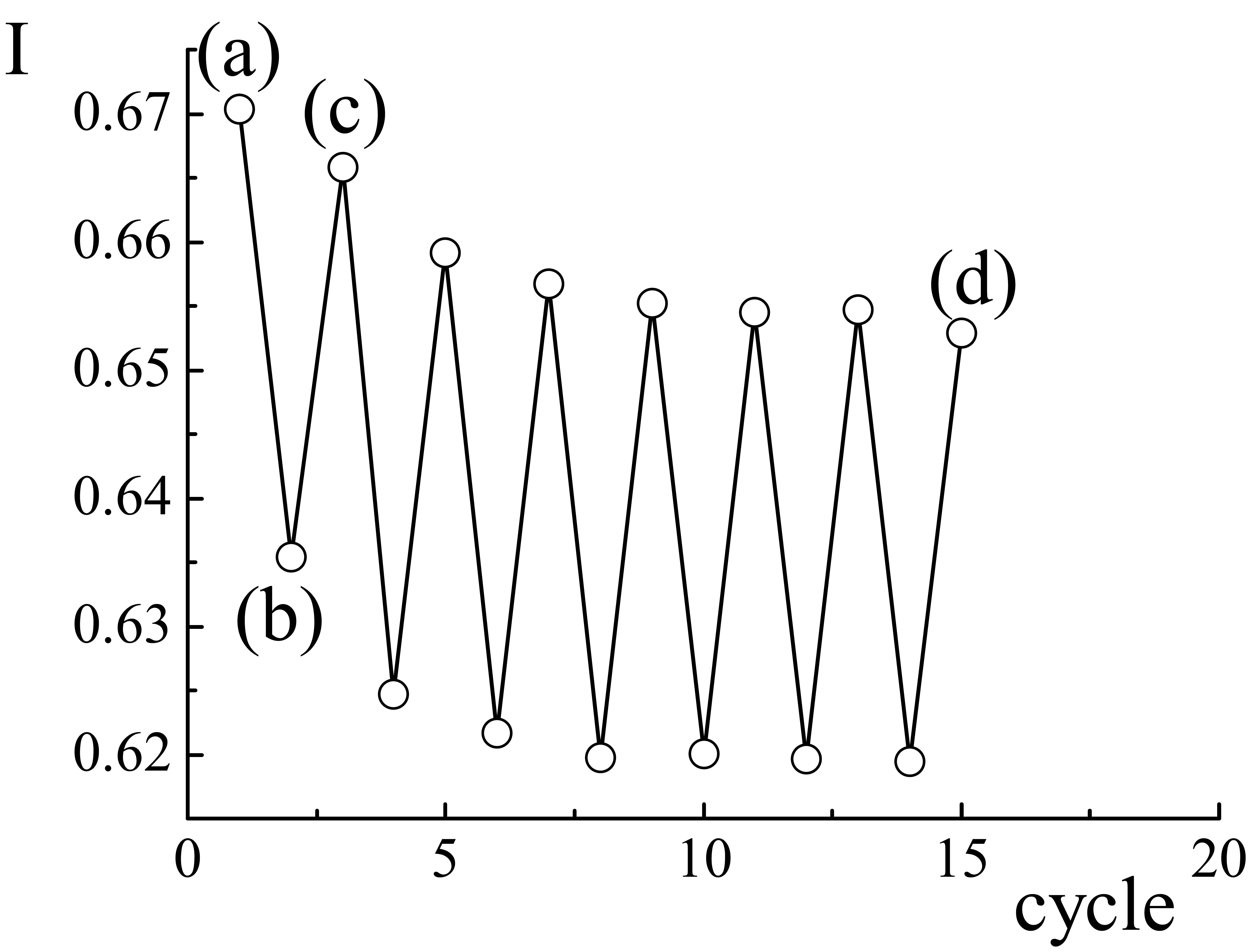}
\caption{The shape of the free bar in Fig. \ref{many_disloc} as a function of successive austenite-martensite transformations. The $I$ parameter reported is proportional to the moment of inertia with respect to $x$, defined as $I=\sum_i \overline{(x_i -\overline{x_i})^2}$.}
\label{ciclos}
\end{figure}

We thus decided to insert by hand an ad hoc distribution of dislocations that produce the TWSME in a sample with free boundaries.
We concatenated many elemental pieces similar to those in Fig. \ref{2disloc}(a), although the separation between dislocations was much shorter.
The sample obtained, after a relaxation step in the austenite phase is shown in Fig. \ref{many_disloc}(a). A detail displaying the distribution of dislocations can be seen in Fig.  \ref{perp}(a).
Note that all dislocations have the same glide directions, and all contribute to a shrinkage along $x$ and expansion along $y$ upon the martensitic transformation. 
The sample after the transition to the martensitic phase is shown in  Fig. \ref{many_disloc}(b), where an appreciable decrease of length is apparent. The sample returns to its initial length after back-transforming to the austenite (Fig. \ref{many_disloc}(c)). The shape change is persistent, as Fig. \ref{ciclos} shows, although we also see in Fig. \ref{many_disloc}(d) 
and \ref{perp}(b) that the sample degrades after cycling. We suspect that the degradation effect would be lesser in larger samples, and also in a three dimensional set up where the movement of dislocations is more limited due to dislocation entanglement.

The present example is a theoretical verification that dislocations can be appropriately distributed in a single crystalline sample in order to induce the TWSME. 

\section{Summary and conclusions}

In this paper we have used a tunable, classical, isotropic potential for identical particles in two dimensions
to study the origin of the two way shape memory effect (TWSME) in martensitic transformations. 
Our model starts with an effective model at intermediate time scales, in which the actual atomic vibrations are not explicitly resolved, but their effect is incorporated in an effective change of the interatomic potential. In this way, subtle effects such as the TWSME can be studied without much numerical effort, something that would not be possible from a full ab initio simulation. 

We first studied the case of a polycrystalline sample, and made clear how the TWSME has its origin in the remaining of martensite pieces of the appropriate orientation inside the austenite, after a cooling-deformation-heating protocol. 
In addition, we were able to show that the same effect is observable in the more natural experimental configuration of a bar, under bending stresses.

In a second part of the work we have seen how the presence of dislocations may induce the TWSME of single crystal samples, as dislocations produce a strain field on its neighborhood that favors the martensitic transformation into variants that produce a stretching of the sample in the insertion plane, and a contraction along the glide direction. We have seen that a properly chosen periodic arrange of dislocations is able to induce a bulk effect with an important shape change.

These findings indicate, first of all, that the present model and simulation techniques are appropriate to discuss subtle effects, as the TWSME, that are well understood on a qualitative basis, but only rather superficially in their microscopical details. It also allows to study problems that are not immediately accessible to present experimental capabilities, as for instance to address the effect of a given dislocation distribution in the shape change of a given sample under martensitic transformation. All these result are promising, and give confidence that other even more demanding studies, as for instance the interplay of martensitic transformation and cracking and fracturing of a sample can also be studied with the kind of model studied here.

\section{Acknowledgments} 

I acknowldege fruitful discussions with G. Bertolino, J. L. Pelegrina, M. Sade and F. Lovey.
This research was financially supported by Consejo Nacional de Investigaciones Cient\'{\i}ficas y T\'ecnicas (CONICET), Argentina. Partial support from grant PICT-2012-3032 (ANPCyT, Argentina) is also acknowledged.

\bibliography{doblem.bib}{}
\bibliographystyle{plain}








\end{document}